\documentclass[5p,twocolumn,authoryear]{elsarticle}

\usepackage{amssymb}

 \usepackage{lineno}
 
 \usepackage{multirow}

\usepackage{graphicx}
\usepackage{dcolumn}
\usepackage{amsmath,amsfonts,amssymb}
\usepackage{bm}
\usepackage{extarrows}  
\usepackage{chemarrow}  
\usepackage{booktabs} 
\usepackage[latin1]{inputenc}
\usepackage[T1]{fontenc}

\usepackage{mathptmx} 

\usepackage{soul} 
\usepackage{color}
\usepackage{soul}

\usepackage{srcltx}

\newcommand{\Fouriertransform}[1]{\ensuremath{\mathcal F}\left\{{#1}\right\}}

\newcommand{\dd}{\ensuremath{\text{d}}}

\newcommand{\taue}{\ensuremath{\tau_\epsilon}}
\newcommand{\taus}{\ensuremath{\tau_\sigma}}

\newcommand{\te}{\tau_{\epsilon}}
\newcommand{\ts}{\tau_{\sigma}}

\def\bfseries{\fontseries \bfdefault \selectfont \boldmath}

\journal{Ultrasound in Medicine and Biology}

\begin{document}
\begin{frontmatter}

\title{Comparison of fractional wave equations for power law attenuation in ultrasound and elastography}

\author[Affil1]{Sverre Holm}
\author[Affil2]{Sven Peter N\"asholm}
\address[Affil1]{Department of Informatics, University of Oslo, P.~O.~Box 1080, NO--0316 Oslo, Norway}
\address[Affil2]{Norsar,  P.~O.~Box 53, NO--2027 Kjeller, Norway}

\date{\today}

\cortext[cor1]{Corresponding Author: Sverre Holm,  P O Box 1080, NO-0316 Oslo, Norway; sverre@ifi.uio.no ; +47 22 85 27 94}

\begin{abstract}
A set of  wave equations with fractional loss operators in time and space are analyzed. It is shown that the fractional Szabo equation, the power law wave equation, and the causal fractional Laplacian wave equation all are low frequency approximations of the fractional Kelvin-Voigt wave equation and the more general fractional Zener wave equation.

The latter two equations are based on fractional constitutive equations while the former wave equations
have been derived from the desire to model power law attenuation in applications like medical ultrasound. This has consequences for use in modelling and simulation especially for applications that do not satisfy the low frequency approximation, such as shear wave elastography. In such applications the wave equations based on constitutive equations are the viable ones.
\end{abstract}


\end{frontmatter}

\pagebreak

\section*{Introduction}

Models for ultrasound attenuation based on relaxation losses have a long history, see e.\ g. 	\cite{Markham1951, bhatia1967ultrasonic}. Waag and his coworkers also contributed with their work on multiple relaxation losses (\cite{Nachman1990}). Similar, but much simpler relaxation models are used for attenuation in salt water and air. In salt water, the two important relaxation processes are due to boric acid and magnesium sulphate (\cite{Ainslie1998}) and in air they are due to nitrogen and oxygen (\cite{Bass1995}). In contrast, in medical ultrasound and elastography, attenuation for both compressional and shear waves often follows a power law  which at first sight is very different from a relaxation model: 
\begin{align}
\alpha_k(\omega) = \alpha_0 \omega^y 
\label{eq:PowerLaw}
\end{align}
where $y$ usually is between 0 and 2 and $\omega$ is angular frequency. The subscript $k$ is used to indicate that this is the imaginary part of the wave number, $k$, and to distinguish it from the order of the fractional derivative, $\alpha$, which will be used later. 

In such media the number of relaxation processes may in practice be uncountable and it has not been possible to make models for attenuation that are so grounded in physical processes as for salt water and air. Nevertheless, the multiple relaxation model with a few processes can be used to model the power-law attenuation as Waag's group and several others have done (\cite{Tabei2003, Yang2005}) over a limited frequency range. But, the relaxation parameters lose their clear physical meaning in this case and become just parameters of a mathematical model. Thus on the one hand there is power law attenuation which is experimentally observed in many different complex materials and on the other hand physical models, exemplified by the multiple relaxation model.

In the last decade or so advances have been made in providing wave equations that partially bridge this gap. They model power law attenuation through the use of fractional, i.e. non-integer, derivatives. To varying degrees these equations are physics-based, but especially for some viscoelastic polymers there is a good physical foundation, \cite{Nasholm2013Zener}. However, there is seldom as direct a connection as for the relaxation models for air and salt water. Nevertheless, they represent one step on the way to a deeper understanding of the underlying physics. They also provide alternative simulations models for wave propagation, often characterized by being parametrized with a low number of parameters. 

For this paper the time fractional derivative is easiest to define in the frequency domain where it is an extension of the Fourier transform of an $n$'th order derivative:
\begin{equation}
\mathcal F\left( \frac{d^\alpha u(t)}{dt^\alpha} \right) = (i\omega)^\alpha U(\omega)
\label{Eq:fractional}
\end{equation}

The fractional derivative of arbitrary order can be understood as a generalization where the integer $n$ 
is replaced with a real number $\alpha$. The close connection between power laws and fractional derivatives is evident from this definition. The time domain equivalent involves a convolution integral implying that the fractional derivative is non-local and has a power-law shaped memory,  \cite{Podlubny1999wholebook}.

The purpose of this paper is to analyze and compare several of the fractional wave equations, to understand their origins, and to relate them to each other. The ultimate objective is to find wave equations which are physically viable. Another more practical objective is to determine which wave equations that are best suited for simulation of medical ultrasound and of shear wave elastography. One of the questions asked is also: What is the most fundamental property to model with a wave equation? Is the objective only to model a power law attenuation, or is it more fundamental to model a medium with viscoelastic properties so that different power law attenuation laws are achieved in different frequency regions? 

Due to the fact that it is usually the low frequency region which is of interest, many wave equations model only this region. But there are applications where the high frequency region is important also, one particularly important one is in shear wave elastography. Even when the interest is only in the low frequency solution, such as for ultrasound imaging, then in order for the solution to be physically viable, it is often an advantage that wave equations give correct solutions beyond this region. The key to obtaining such results is the viscoelastic constitutive equation. The paper therefore builds a case for the viewpoint that it is the viscoelastic constitutive equation which is the more fundamental and physical property, rather than the power law characteristics.

It should be noted that low frequencies here means low in comparison to the value of a time constant ($\taus$ in e.g.\ Eq.\ (\ref{eq:WaveCaputo})). Thus for e.g. compressional waves in medical ultrasound, low frequencies stretch well beyond the usual range and up into the 100's of MHz or higher. For such an application the power law has an exponent, $y$, between one and two. On the other hand, it turns out that for shear waves in the human body in elastography, the low frequency limit is at less than 10 Hz (\cite{Holm2010}). Thus elastography operates above the low frequency limit with $y$ less than unity.

\subsection*{Physcially viable solutions}

There are some criteria that need to be satisfied by the solutions of wave equations in order for them to be physically viable. The first is that of causality: No output is allowed before there is an input.

The second property is that the loss function cannot rise arbitrarily fast and has to satisfy:
\begin{align}
\lim_{\omega\to\infty} \alpha_k(\omega) / \omega =  0 
\label{eq:AsymptoticCondition}
\end{align}

This means that the exponent, $y$, has to be less or equal to one in the high frequency limit. This condition was first given by \cite{Weaver1981} and more rigorously by \cite{Hanyga2013Wave}. It follows from causality, passivity, and linearity and is a consequence of the property that viscoelasticity can be modelled as a sum of many spring-dashpot mechanical models with realistic, i.e. positive, constants, \cite{beris1993admissibility}.

\section*{Fractional wave equations}

\begin{figure}[hbt]
  \centering
  \includegraphics[width=10cm]{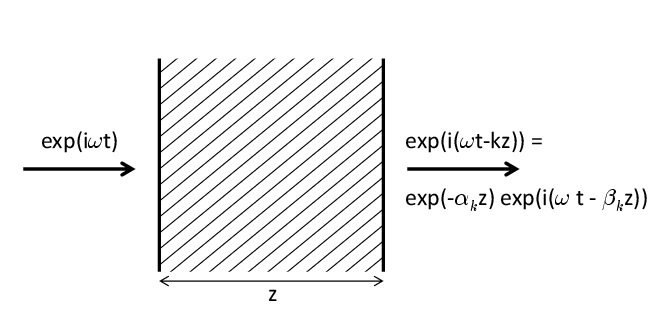}
  \caption{A slab of thickness $z$ with a plane wave travelling from left to right.}
  \label{fig:slab}
\end{figure}

We will give wave equations for plane waves in a homogenous medium. Fig.\ \ref{fig:slab} illustrates how the wave is modified by transmission through a slab of thickness $z$. The wave number is in general complex and given by $k = \beta_k-i\alpha_k $ where the real part describes dispersion and the imaginary part gives the attenuation. Note that many papers use the opposite sign for the exponent in the travelling wave. This will result in somewhat different wave equations. This is shown in the Appendix A which has been included in order to facilitate comparisons between papers using different sign conventions.

\subsection*{Fractional time loss operator, type 1}
One of the oldest fractional wave equations is due to \cite{Caputo1967}. In the terminology of \cite{Wismer06} it is
\begin{align}
\nabla^2 u -\dfrac 1{c_0^2}\frac{\partial^2 u}{\partial t^2} + \taus^{y-1} \dfrac{\partial^{y-1}}{\partial t^{y-1}}\nabla^2 u  = 0,
\label{eq:WaveCaputo}
\end{align}
where $u$ is the displacement, $c_0$ is the phase velocity at zero frequency, $\alpha_0 = \taus/(2c_0)$ in Eq. (\ref{eq:PowerLaw}), and the condition $(\omega  \taus)^{y-1} \ll 1$ has to be satisfied. Here we will call it \emph{the fractional Kelvin-Voigt wave equation}. This equation and all subsequent wave equations in this paper could equally well have been formulated for the particle velocity, $v=\dot{u}$, and in the case of compressional waves also the pressure, $p$. 

For low frequencies relative to the relaxation time constant $\taus$, 
\begin{align}
(\omega  \taus)^\alpha \ll 1, 
\label{eq:lowfreqCond}
\end{align}
$\alpha$ and the power law exponent of Eq.\ (\ref{eq:PowerLaw}) are related by $y = \alpha+1$ where $0<\alpha\le1$.

\emph{The fractional Zener wave equation} derived by \cite{Holm2011} is an extension of Eq.\ (\ref{eq:WaveCaputo}) with an additional loss term which contributes to a different behavior for very high frequencies:
\begin{align}
\nabla^2 u -\dfrac 1{c_0^2}\frac{\partial^2 u}{\partial t^2} + \taus^\alpha \dfrac{\partial^\alpha}{\partial t^\alpha}\nabla^2 u	- \dfrac {\taue^\beta}{c_0^2} \dfrac{\partial^{\beta+2} u}{\partial t^{\beta+2}} = 0.
\label{eq:ZenerWave}
\end{align}

These equations are here said to have a fractional time loss operator of type 1, because there is a term with a mix of a fractional time derivative and a non-fractional Laplacian in the loss term. 
\subsection*{Fractional time loss operator, type 2}

Type 2 of the fractional time loss operator is an operator which has only temporal derivatives. The distinction between types 1 and 2 is only to aid in the presentation of the various loss operators. The first type 2 operator is the general Szabo convolution model:
\begin{align}
\nabla^2 u  - \frac{1}{c_0^2} \frac{\partial^2 u}{\partial t^2}
- L_y(t) *  u = 0,
\label{eq:Szabo}
\end{align}
where $* $ is the temporal convolution operator. The causal loss operator, $L_y(t)$ is given by the power law attenuation that it describes and has different forms for even and odd integer values of $y$ and for the non-integer case, see \cite{Szabo94, szabo2004diagnostic} for details.
This wave equation can be written in terms of fractional derivatives (\cite{Chen03}) and in the terminology of \cite{zhang2012modified} \emph{the fractional Szabo equation} is:
\begin{align}
\nabla^2 u  - \frac{1}{c_0^2} \frac{\partial^2 u}{\partial t^2}  
- \frac{2 \alpha_0}{\cos(\pi y/2)} \frac{\partial^{y+1}u}{\partial t^{y+1}} 
= 0.
\label{eq:SzaboFractional}
\end{align}

\cite{Kelly2008} added an extra loss term to Eq.\ (\ref{eq:SzaboFractional}) in order to  model power law losses with exactly the same exponent, $y$, for all frequencies. Therefore we will call it \emph{the power law wave equation}:
\begin{align}
\nabla^2 u  - \frac{1}{c_0^2} \frac{\partial^2 u}{\partial t^2}  
- \frac{2 \alpha_0}{c_0 \cos(\pi y/2)} \frac{\partial^{y+1}u}{\partial t^{y+1}} 
- \frac{\alpha_0^2}{\cos^2(\pi y/2)} \frac{\partial^{2y}u}{\partial t^{2y}} 
= 0.
\label{eq:KellyWaveEq}
\end{align}

\subsection*{Fractional Laplacian loss operator}

\cite{Chen04} proposed \emph{the fractional Laplacian wave equation}:
\begin{align}
\nabla^2 u  - \frac{1}{c_0^2} \frac{\partial^2 u}{\partial t^2} 
- 2 \alpha_0 c_0^{y-1}\frac{\partial}{\partial t}  (-\nabla^2)^{y/2} u = 0.
\label{eq:fractLaplacian}
\end{align}

It was amended by the addition of an extra loss term by \cite{Treeby2010}:
\begin{align}
\nabla^2 u  - \frac{1}{c_0^2} \frac{\partial^2 u}{\partial t^2} 
+ \tau \frac{\partial}{\partial t} (-\nabla^2)^{y/2} u
+ \eta (-\nabla^2)^{(y+1)/2} u = 0,
\label{eq:TreebyCoxWave}
\end{align}
where $\tau = -2\alpha_0 c_0^{y-1}$ and $\eta=2 \alpha_0 c_0^y \tan(\pi y/2)$.  The additional loss term makes the wave equation causal at low frequencies, so a proper name is \emph{the causal fractional Laplacian wave equation}. Note however that the sign of this last term will depend on the definition of the Fourier transform used in the derivation, as explained in Appendix A.
%

\subsection*{Convolution loss operator}
The most general way to express the loss term in the wave equations is by means of a convolution operator. It can model losses which are different from power laws also. The convolution terminology is taken from Sec.\ (4.2.2) in \cite{Mainardi2010} and in the creep representation it is:
\begin{align}
\nabla^2 u  - \frac{1}{c_0^2} \frac{\partial^2 u}{\partial t^2}  
- \frac{1}{c_0^2}\big( \Psi(t) * \big) \frac{\partial^2 u}{\partial t^2} = 0,
\label{eq:Creep}
\end{align}
where $\Psi(t) = [dJ(t)/dt]/J_g$ is the rate of creep, $J(t)$ is the creep compliance, and $J_g=J(0^+)$ is the glass compliance.  The various fractional operators discussed in this paper are such that they result in power law attenuation. Thus by comparison with the fractional Szabo equation, Eq.\ (\ref{eq:SzaboFractional}), the convolution with the rate of creep corresponds to the operator:
\begin{align}
\big( \Psi(t) * \big) = \frac{2\alpha_0 c_0^2}{cos(\pi y/2)} \frac{\partial^{y-1}}{\partial t^{y-1}}.
\end{align}

The alternative relaxation representation of Eq.\ \eqref{eq:Creep} is, \cite{Mainardi2010}:
\begin{align}
\nabla^2 u  - \frac{1}{c_0^2} \frac{\partial^2 u}{\partial t^2}  
+ \big( \Phi(t) * \big)  \nabla^2 u = 0.
\label{eq:Creep2}
\end{align}

where $\Phi(t) = [dG(t)/dt]/G_g$ is the rate of relaxation, $G(t)$ is the relaxation modulus, and $G_g =G(0^+)$ is the glass modulus.
Hence the fractional Kelvin-Voigt wave equation, Eq.\ (\ref{eq:WaveCaputo}), can be formulated by using:
\begin{align}
\big( \Phi(t) * \big) = \taus^\alpha \frac{\partial^\alpha}{\partial t^\alpha}.
\end{align}

Towards the end of the paper, we will also formulate the Nachman-Smith-Waag model in the convolution framework.

\section*{Wave equations derived from viscoelastic models}
The most general constitutive equation considered  here is the fractional Zener model, or the fractional standard linear solid. Following \cite{Holm2011} it is:
\begin{align}
	\sigma(t) +\tau_{\epsilon}^{\beta} \frac{\partial^{\beta}\sigma(t)}{\partial t^{\beta}}  = G_e \left[\epsilon(t) +\tau_{\sigma}^{\alpha} \frac{\partial^{\alpha}\epsilon(t)}{\partial t^{\alpha}}\right], 
\label{Eq:gZener}
\end{align}
where $\sigma(t)$ is stress, $\epsilon(t)$ is strain, and the elastic modulus is the equilibrium modulus. In the general case, stress and strain are tensors, but here it is sufficient to represent them as scalars in order to demonstrate the properties of the resulting wave equations.  This model is also a special case of a more general convolution model between stress and strain expressed by the creep compliance,  $J(t)$, or the relaxation modulus, $G(t)$.

This constitutive equation is well founded in material science as a good description of material properties as reviewed in Ch.\ 3 of \cite{Mainardi2010} and the survey paper \cite{Nasholm2013Zener}. There the conditions $\tau_{\sigma}^\alpha \ge \tau_{\epsilon}^\beta$ and $\alpha \le \beta$ are justified also. It is also common to assume the even stricter condition that $\alpha = \beta$ which will be done in the rest of this paper also. This condition is necessary in order to make the system thermodynamically well-behaved. As an example, the model has been used to model shear wave propagation in human tissue in elastography, \cite{Klatt2007}. 

Although the constitutive equation gives a good description of materials, our understanding of why this is so is still lacking, i.e. what is it in the underlying microstructure of the materials that gives this behavior? Therefore the fractional parameter $\alpha$ cannot be determined directly from a physical model, but it is closely related to the power law behavior of the attenuation see belowl. One of the advantages of the model over the multiple relaxation model (\cite{Nachman1990}) is parsimony, i.\ e.\ that it is described by only four parameters ($G_e$, $\tau_{\sigma}$, $\tau_{\epsilon}$, and $\alpha$). This is fewer than the number of parameters required for the multiple relaxation model except for the very simplest of media such as air and salt water. This is usually an advantage in simulations.

When Eq.\ (\ref{Eq:gZener}) is combined with conservation of mass and momentum, it leads to the fractional Zener wave equation, Eq. (\ref{eq:ZenerWave}), for which
%
%
the dispersion relation is:
\begin{align}
k^2 - \frac{\omega^2}{c^2_0}  + (\tau_{\sigma}i \omega)^{\alpha} k^2  - (\tau_{\epsilon}i \omega)^{\beta} \frac{\omega^2}{c^2_0} = 0
\label{Eq:dispersionZener}
\end{align}

The simpler, three-term Kelvin-Voigt constitutive equation is obtained from Eq.\ (\ref{Eq:gZener}) by setting $\tau_{\epsilon}=0$. Examples of fitting of that model to shear wave elastography data can be found in \cite{zhang2007congruence}. The fractional Kelvin-Voigt constitutive equation leads to the wave equation first derived by Caputo, Eq.\ (\ref{eq:WaveCaputo}).

It should be noted that there are several ways that this result can be extended for the special case of pressure waves in fluids and liquids. First, the model may include nonlinear acoustics as done in \cite{Prieur2011, Prieur2012}. Second, in a complete acoustic theory for fluids and liquids a thermal loss term may also be significant. This was analyzed in the former two references as well as summarized in \cite{Holm2013deriving}.
 
This paper will build on the asymptotic results previously determined for the low, intermediate and high frequency regions  of Eq.\ (\ref{eq:ZenerWave}). For completeness the main results from \cite{Holm2011} are  repeated here. The attenuation follows these asymptotes:
\begin{align}
\alpha_k(\omega) \propto 
   \begin{cases}
    \omega^{1+\alpha},         & (\omega \tau_\epsilon)^{\alpha} \le (\omega \tau_\sigma)^{\alpha} \ll 1 \\
    \omega^{1-\alpha/2},       & (\omega\te)^\alpha \ll 1 \ll  (\omega\ts)^\alpha\\
    \omega^{1-\alpha},          &  1 \ll (\omega \tau_\epsilon)^{\alpha} \le (\omega \tau_\sigma)^{\alpha}
   \end{cases}
\label{Eq:attenuation3regimes}
\end{align}
The phase velocity also has three regimes:
\begin{align}
c_p(\omega) =
   \begin{cases}
    c_0,				& (\omega \tau_\epsilon)^{\alpha} \le  (\omega \tau_\sigma)^{\alpha} \ll 1 \\
    \propto  \omega^{\alpha/2},       & (\omega\te)^\alpha \ll 1 \ll  (\omega\ts)^\alpha\\
    c_0 \left( {\tau_{\sigma}}/{\tau_{\epsilon}}   \right)^{\alpha/2},   &  1 \ll (\omega \tau_\epsilon)^{\alpha} \le (\omega \tau_\sigma)^{\alpha}
   \end{cases}
\label{Eq:cp3regimes}
\end{align}

In the case of the fractional Kelvin-Voigt model with $\tau_{\epsilon}=0$, the third regime will disappear for both the attenuation and the phase velocity. In that case one can recognize the well-known result for the viscous loss model with a non-fractional derivative, $\alpha=1$, (Stokes' equation) which gives attenuation proportional to $\omega^2$ for low frequencies and proportional to $\sqrt{\omega}$ for high frequencies.

The following sections builds on a set of approximations derived for the low-frequency regime which are given in Appendix B. They will be used subsequently to find various interesting approximations to the lossy wave equation. 

\section*{Wave equations with fractional temporal derivative loss operator}
An approximation 
which is valid for low frequencies is obtained by replacing $k^2$ by $(\omega/c_0)^2$, Eq.\ (\ref{Eq:lowlowFreq}), in the viscous terms of the dispersion equation for the fractional Kelvin-Voigt equation (Eq. (\ref{Eq:dispersionZener}), with $\tau_{\epsilon} = 0$):
\begin{align}
k^2 - \frac{\omega^2}{c^2_0}  + (\tau_{\sigma}i \omega)^{\alpha} (\omega/c_0)^2  = 0.
\end{align}
This leads to:
\begin{align}
\nabla^2 u  - \frac{1}{c_0^2} \frac{\partial^2 u}{\partial t^2}  
+ \frac{\tau_\sigma^\alpha}{c_0^2} \frac{\partial^{\alpha+2}u}{\partial t^{\alpha+2}}
= 0.
\label{eq:Szabofract2}
\end{align}
This is the Szabo wave equation written in terms of fractional derivatives as given in Eq.\ (\ref{eq:SzaboFractional}). Thus this is one alternative for a low frequency variant of Eqs. (\ref{eq:WaveCaputo}) and  (\ref{eq:ZenerWave}). Because it can be derived directly from the causal Kelvin-Voigt wave equation, it is causal for low frequencies as demonstrated in e.g. \cite{zhang2012modified} also. Here it has been derived as a low frequency approximation to Eq.\ (\ref{eq:WaveCaputo}) and this is fully consistent with the low frequency approximation which is a central step in Szabo's own derivation. 

If one for a moment neglects that it is based on a low frequency assumption, then it is straightforward to show that for high frequencies, the attenuation will be proportional to $\omega^{1+\alpha/2}, ~ 0<\alpha\le 1$. But at these frequencies, the equation is used for something which violates its assumptions. It is therefore not surprising that this also conflicts with the condition for a physically viable solution given by Eq. (\ref{eq:AsymptoticCondition}). 

\subsection*{Two loss terms}
In the derivation above,  the starting point was the Kelvin-Voigt wave equation or its generalization the Zener wave equation. Then low frequencies were assumed according to Eq.\ (\ref{eq:lowfreqCond}) so that the time-space substitution $k \approx \omega/c_0$ could be used. An alternative is to say that the power law attenuation and the accompanying dispersion required from causality is the goal. The objective is then to find a wave equation which satisfies power law attenuation exactly. This is the starting point of \cite{Kelly2008} and it is the same as requiring that the low frequency approximation for the wave number in Eq.\ (\ref{Eq:lowFreq3}) should be satisfied exactly for all frequencies, not just for low frequencies. That equation is the same as Eq.\ (7) in \cite{Kelly2008} when the difference in sign conventions for the Fourier transform have been taken into account (see Appendix A).

Squaring both sides of  Eq.\ (\ref{Eq:lowFreq3}) gives this dispersion relation:
\begin{align}
k^2  =  \left[ \frac{\omega}{c_0} - \frac{\alpha_0 i^{y+1} \omega^y}{\cos{\pi y/2}}\right]^2
=  \frac{\omega^2}{c_0^2} - \frac{2 \alpha_0 }{c_0 \cos(\pi y/2)}  (i\omega)^{y+1} 
\notag\\
- \frac{\alpha_0^2}{\cos^2(\pi y/2)}(i\omega)^{2y}
\label{eq:Kellydispersion}
\end{align}

Inverse Fourier transformation in space and time yields Eq.\ (\ref{eq:KellyWaveEq}) which has an extra loss contribution compared to Eq.\ (\ref{eq:Szabofract2}). It is repeated here:
\begin{align}
\nabla^2 u  - \frac{1}{c_0^2} \frac{\partial^2 u}{\partial t^2}  
- \frac{2 \alpha_0}{c_0 \cos(\pi y/2)} \frac{\partial^{y+1}u}{\partial t^{y+1}} 
- \frac{\alpha_0^2}{\cos^2(\pi y/2)} \frac{\partial^{2y}u}{\partial t^{2y}} 
= 0
\label{eq:KellyWaveEq2}
\end{align}

\cite{Kelly2008} say that the last term usually can be neglected since it is proportional to $\alpha_0^2$ and thus is small compared to the main loss term. Therefore it doesn't make much difference. This is certainly the case for low frequencies according to the condition in Eq.\ (\ref{eq:lowfreqCond}), where $\alpha_0$ is proportional to $\taus$.

But for high frequencies, the enforced power law relation of Eq.\ (\ref{Eq:lowFreq3}) still holds exactly, and does so for however high frequencies one tries for. As the frequency increases above the low frequency condition of Eq.\ (\ref{eq:lowfreqCond}), the last loss term will play a larger and larger role. This wave equation therefore models the same power law attenuation for all frequencies. Most other wave equations only model this attenuation for low frequencies.


But, as long as the exponent, $y$, is larger than 1, the condition for a physically viable solution according to Eq.\ (\ref{eq:AsymptoticCondition}) is violated. This is the limitation of the wave equation of \cite{Kelly2008}: At low frequencies the extra loss term can be neglected, and at high frequencies it may make the solution physically unviable. This problem was also illustrated by \cite{johnsonnumerical} through numerical simulations.


The paper by \cite{kowar2011causality} takes the same approach as \cite{Kelly2008} and tries to model a power law attenuation exactly. As a result their Eq.\ (36) is similar to Eq.\ (\ref{eq:Kellydispersion}). However, \cite{kowar2011causality} go further  as they have observed that the result is non-causal when $y>1$. They then amend their wave equation in order to make it causal. The result, Eq.\ (61) in  \cite{kowar2011causality} is a wave equation which after some reorganization can be written on this form:
\begin{align}
	\nabla^2 p  
	& - \dfrac{\alpha_1^2+1}{c_\infty^2} \frac{\partial^2 p}{\partial t^2} 
          +\tau_0^{y-1}\dfrac{\partial^{y-1}}{\partial t^{y-1}}\nabla^2 p
           -\dfrac{\tau_1^{y-1}}{c_\infty^2} \dfrac{\partial^{y+1}p}{\partial t^{y+1}}\notag\\
	&- \dfrac{2\alpha_1}{ c_\infty^2} \Fouriertransform{\sqrt{1+(i\omega\tau_0)^{y-1}}} * \frac{\partial^2 p}{\partial t^2}
	= 0
\label{eq:KowarWaveEq}
\end{align}
where $c_\infty = c_0(1+\alpha_1)$ is the asymptote of the phase velocity at infinite frequency. The first four terms are similar to those of the fractional Zener wave equation, Eq.\ (\ref{eq:ZenerWave}). The final Fourier transform term can be approximated for different frequency regions and it can be shown that it will merge into the second and fourth terms.  In our view \cite{kowar2011causality} give an argument, albeit a rather indirect and involved one, for why the fractional Zener wave equation has the desirable property of physical validity for all frequencies.

\section*{Wave equations with  a fractional Laplacian loss operator}
In this section we go back to Eq.\ (\ref{Eq:dispersionLow}) and use substitutions based on $k \approx \omega/c_0$ in the loss terms as in the previous section. The difference is that we only partially substitute time for space. By assuming low frequencies and replacing $\omega^\alpha$ by $\omega\cdot (c_0k)^{\alpha-1}$ in the first loss term and by using  $(c_0k)^\alpha$ instead of $\omega^\alpha$ in the second loss term one gets:
\begin{align}
k^2 - \frac{\omega^2}{c^2_0}
+  i \sin(\pi\alpha/2) \tau_{\sigma}^\alpha c_0^{\alpha-1} \omega k^{\alpha+1} 
\notag\\
+ \cos(\pi\alpha/2) \tau_{\sigma}^\alpha c_0^{\alpha} k^{\alpha+2} = 0
\label{Eq:dispersionTreebyCox2010A}
\end{align}
The equivalent wave equation using the fractional Laplacian introduced by \cite{Chen04} is:
\begin{align}
\nabla^2 u  - \frac{1}{c_0^2} \frac{\partial^2 u}{\partial t^2}
- \sin(\pi\alpha/2) \tau_{\sigma}^\alpha c_0^{\alpha-1} \frac{\partial}{\partial t} (-\nabla^2)^{(\alpha+1)/2} u&\notag\\
+ \cos(\pi\alpha/2) \tau_{\sigma}^\alpha c_0^{\alpha} (-\nabla^2)^{\alpha/2+1} u &= 0
\label{eq:fractLaplacianCausal}
\end{align}

First let's neglect the last term. As $ \sin(\pi\alpha/2) \tau_{\sigma}^\alpha c_0^{\alpha-1} = 2  \alpha_0 c_0^\alpha =  2  \alpha_0 c_0^{y-1}$ using Eq.\ (\ref{Eq:low-alpha}) and $y=\alpha+1$, one then obtains the lossy wave equation, Eq.\ (\ref{eq:fractLaplacian}), with a fractional Laplacian attenuation operator first proposed by \cite{Chen04}. Due to the neglect of the second term, this equation does not satisfy the Kramers-Kronig condition as has been pointed out by several researchers. The fractional Laplacian wave equation of Eq.\ (\ref{eq:fractLaplacian}) is therefore a non-causal low frequency approximation to the fractional Kelvin-Voigt wave equation of Eq.\ (\ref{eq:WaveCaputo}).

By keeping all terms in Eq.\ (\ref{eq:fractLaplacianCausal}), and using the substitutions of the previous paragraph, one gets:
\begin{align}
\nabla^2 u  - \frac{1}{c_0^2} \frac{\partial^2 u}{\partial t^2}
- 2  \alpha_0 c_0^{y-1} \frac{\partial}{\partial t} (-\nabla^2)^{(\alpha+1)/2} u&\notag\\
-2\alpha_0c_0^y \tan(\pi y/2) (-\nabla^2)^{\alpha/2+1} u &= 0
\label{eq:fractLaplacianCausal2}
\end{align}
This is almost Eq.\ (\ref{eq:TreebyCoxWave}), the equation derived by \cite{Treeby2010} to be causal. There is however one minor difference and that is the sign of the last term. This is caused by the differences in the sign of the Fourier transform kernel between \cite{Treeby2010} and this paper, see the Appendix A. This wave equation is therefore a causal low frequency approximation to the fractional Kelvin-Voigt wave equation of Eq.\ (\ref{eq:WaveCaputo}) expressed with  fractional Laplacians instead of time-fractional operators. Unlike for the Szabo wave equation, it is not possible to find an analytic solution of the fractional Laplacian wave equation beyond the region of validity and into the high frequencies, and in that way show that it violates Eq.\ (\ref{eq:AsymptoticCondition}). 

The substitution of time with space is entirely due to the implicit low-frequency approximation and the use of Eq.\ (\ref{Eq:lowlowFreq}). Unfortunately we doubt that there is much significance beyond that of the fractional Laplacian, i.e. little or no relationship to the spatial structure of the medium.

\section*{Convolution formulation of the multiple relaxation model}

Finally we would like to point out some connections between fractional models and the Nachman-Smith-Waag multiple relaxation model (\cite{Nachman1990}).
As shown in Section 4.2 of \cite{Nasholm2013Zener}, conservation of mass and conservation of momentum give the general dispersion relation
\begin{align}
	k^2(\omega) = \omega^2\rho_0\kappa(\omega),
	\label{eq:dispersion_relation}
\end{align}
where the frequency-domain generalized compressibility is $\kappa(\omega) \equiv \epsilon(\omega)/\sigma(\omega)$ and therefore given directly by a constitutive equation such as Eq.\ (\ref{Eq:gZener}). This relation can be inverse Fourier transformed into 
\begin{align}
	\nabla^2 u(x,t) = \rho_0 \kappa(t) * \dfrac{\partial^2 u(x,t)}{\partial t^2},
	\label{eq:Conv}
\end{align}
where $\kappa(t)$ is the temporal representation of the generalized compressibility. Notice that the formulation of Eq.~(\ref{eq:Conv}) is related to the wave equation (1) in \cite{Verweij2013} (see also the references therein). These two wave equations are mathematically similar if nonlinear terms are disregarded, and if $m(t)$ in \cite{Verweij2013} is identified as $m(t) = \rho_0c_0^2 \kappa(t)$. Related similarity is observed between Eq.~(37) and the random effective propagation model Eq.~(16) in \cite{Garnier2010fractionalprecursors}. %

\cite{Nasholm2011} considered the integral generalization
\begin{align}
	\kappa(\omega) = \kappa_0-i\omega\int_0^\infty \dfrac{\kappa_\nu(\Omega)}{\Omega+ i\omega} \dd \Omega
	\label{eq:NSW_integral_form}
\end{align}
of the Nachman--Smith--Waag generalized compressibility, ~\cite{Nachman1990}
\begin{align}
	\kappa(\omega) = \kappa_0-i\omega \sum_{\nu=1}^N \dfrac{\kappa_\nu\tau_\nu}{1 + i\omega\tau_\nu},
\end{align}
where the set of $\tau_\nu$ are the relaxation time constants and $\kappa_\nu$ are the corresponding discrete compressibility contributions. For the continuous distribution of compressibility contributions of Eq.\ (\ref{eq:NSW_integral_form}), $\kappa_\nu(\Omega)$ denotes the contribution at relaxation frequency $\Omega$.
Now combining Eq.~\eqref{eq:NSW_integral_form} with Eq.~\eqref{eq:dispersion_relation} and taking the inverse Fourier transform gives the convolution form of the multiple relaxation wave equation:
\begin{align}
	\nabla^2 u -\dfrac{1}{c_0^2}\dfrac{\partial^2 u}{\partial t^2} %
		-\dfrac{\partial}{\partial t} \left( %
		\int_0^\infty H(t)e^{-\Omega t}\kappa_\nu(\Omega) \dd \Omega \right) 
			* \dfrac{\partial^2 u}{\partial t^2} = 0,
	\label{eq:NSW_we_1}
\end{align}
where $H(t)$ is the Heaviside step function and the relation $c_0^2 = 1/(\rho_0\kappa_0)$ is applied. This is a special case of Eq.\ (\ref{eq:Creep}) and now we want to find the convolution kernel, the rate of the creep. Using the arguments of \cite{Nasholm2013JASA}, Eq.~\eqref{eq:NSW_we_1} can be rewritten in the form
\begin{align}
	\nabla^2 u -\dfrac{1}{c_0^2}\dfrac{\dd^2 u}{\dd t^2} %
		-\dfrac{\partial}{\partial t} %
		H(t) \mathcal L \left\{ \kappa_\nu(\Omega)\right\}   
			* \dfrac{\partial^2 u}{\partial t^2} = 0,
	\label{eq:NSW_we_2}
\end{align}
where $\mathcal L$ signifies the Laplace transform from the $\Omega$ domain to the $t$ domain. Thus the multiple relaxation wave equation can be expressed in the creep representation form of Eq.\ (\ref{eq:Creep}) with %
\begin{align}
\Psi(t) = c_0^2 \dfrac{\partial}{\partial t} H(t) \mathcal L \left\{ \kappa_\nu(\Omega)\right\} 
\label{eq:NSW-creep}
\end{align}
This is an equivalent, but often straightforward form of the wave equation than that given in \cite{Nachman1990}. In addition, we note that Eq.\ (\ref{eq:NSW_we_2}), the convolution representation of the Nachman-Smith-Waag wave equation, is similar to the Szabo model, Eq.\ \eqref{eq:Szabo}, so the Szabo kernel can also be found. Other parallels can also be drawn such as the relationship between the fractional Kelvin-Voigt and Zener wave equations. This has been discussed in \cite{Nasholm2011} and \cite{Nasholm2013JASA}.

\section*{Conclusion}

Four of the common wave equations with time- and space-fractional loss operators, here denoted as the fractional Szabo equation, the power law wave equation, the fractional Laplacian wave equation, and the causal fractional Laplacian wave equation (Eqs.\ (\ref{eq:SzaboFractional}) - (\ref{eq:TreebyCoxWave})), have been analyzed. It has been shown that they can all be derived as low frequency approximations to {the fractional Kelvin-Voigt wave equation} which again is a special case of the more general {fractional Zener wave equation}, Eqs (\ref{eq:WaveCaputo}) and (\ref{eq:ZenerWave}).


The latter two wave equations are based on fractional constitutive equations which are fairly well established as descriptions of realistic materials, while the former four wave equations were derived so that they produce the power law solutions which are often observed in measurements. The fractional Kelvin-Voigt and the fractional Zener wave equations therefore possess more desirable physical properties than the others. 

The fractional Kelvin-Voigt and  Zener wave equations can also be viewed as special cases of wave equations with convolution kernels for loss terms. Moreover, Eq.\ (\ref{eq:NSW_we_2}), provides a straightforward convolution form of the Nachman-Smith-Waag wave equation which is linked to the creep representation of Eq.\ (\ref{eq:Creep}). What is so attractive about using fractional derivatives rather than general convolution kernels, is that they provide solutions that fit well with the power laws that are frequently observed.

For use in numerical evaluation where the excitation source and medium of propagation satisfy the low frequency approximation, such as in medical ultrasound, there is a wide choice of equations to use, and one can use either one of the exact or low-frequency approximations given here.  The choice may be dictated by numerical considerations which are beyond the scope of this paper.  

It should however be pointed out that for simulation of waves that do not satisfy the low frequency approximation, and shear waves in elastography is an important example of that, these low frequency approximations cannot be used. Here either the fractional Kelvin-Voigt or the fractional Zener equations are the only ones which describe the properties of the wave propagation and they are also the best ones for simulations. 

\section*{Acknowledgement}

We want to thank the reviewers for comments that helped improve the presentation.

\appendix
\section{Sign convention in Fourier transform}

During the course of this study it was found that the two sign conventions that are possible for the Fourier transform affect the results of various papers and make comparisons difficult. Therefore the consequences of the two sign conventions are listed here, along with the key papers that use either one of them.

\scriptsize

\begin{center}
   \begin{tabular}{ l || l | l }
     \hline
Property & Convention 1 & Convention 2\\ \hline \hline
Fourier transform				& $\int u(x,t) \exp(i(kx - \omega t))dxdt$&  $\int u(x,t) \exp(i( \omega t - kx))dxdt$ \\ \hline
$\partial^\eta u_1/\partial t^\eta$ 	& $(i\omega)^\eta  u_1$ 			&  $(-i\omega)^\eta  u_2$ \\ \hline
$\partial^\eta u_1/\partial x^\eta$	& $(-ik)^\eta u_1$ 				&  $(ik)^\eta u_2$ \\ \hline 
$(-\nabla^2)^{\eta/2} $			&  $-|k|^{\eta}$ 					& $-|k|^{\eta}$ \\ \hline\hline

Plane wave 					& $u_1(x,t) = \exp(i(\omega t - kx))$ 	&  $u_2(x,t) = \exp(i(kx - \omega t))$ \\ \hline
Wave number				&  $k = \beta_k - i\alpha_k$		&  $k = \beta_k + i\alpha_k$ \\ \hline
Attenuated wave				& $\exp(-\alpha_k x) \exp(i(\omega t - \beta_kx))$  & $\exp(-\alpha_k x) \exp(i(\beta_kx - \omega t))$ \\ \hline \hline
Used by					& \cite{Wismer06}				& \cite{Nachman1990}\\ 
						& \cite{Holm2010}				& \cite{Szabo94} \\
						& \cite{Mainardi2010}			& \cite{Podlubny1999wholebook}\\ 
						& \cite{Caputo2011} 			& \cite{Chen03}  \\		
						& \cite{Holm2011}	           		& \cite{Chen04}  \\ 
						& \cite{Nasholm2011}			&  \cite{Kelly2008}\\ 
						& \cite{zhang2012modified}		&  \cite{Treeby2010}\\
                                                                & This paper                                            & \cite{kowar2011causality}\\
\hline

   \end{tabular}
\end{center}

\normalsize

\section{Approximations in the low-frequency regime}
Various approximations based on the fractional Zener and the Kelvin-Voigt wave equations are derived in order to show how they relate to the wave equations of Eqs.\ (\ref{eq:SzaboFractional}) - (\ref{eq:TreebyCoxWave}). The approximations are all based on the the low-frequency region, the first regime in Eqs. (\ref{Eq:attenuation3regimes}-\ref{Eq:cp3regimes}),  where the frequency is low compared to both time constants. The first result needed for later derivations can be found by solving Eq.\ (\ref{Eq:dispersionZener}) approximately for $k$: 
\begin{align}
k = \frac{[1+(\tau_\epsilon i \omega)^\alpha)]^{0.5}} {[1+(\tau_\sigma i \omega)^\alpha)]^{0.5}}
\approx  \frac{\omega}{c_0} \left[1 - \frac{1}{2}( \tau_{\sigma}^{\alpha}- \tau_{\epsilon}^{\alpha}) ( i \omega)^{\alpha}\right].
\label{Eq:lowFreq}
\end{align}
Because $\tau_{\sigma} \ge \tau_{\epsilon}$ one can set $\tau_{\epsilon} = 0$ in this frequency regime without loss of generality or
\begin{align}
k \approx  \frac{\omega}{c_0} \left[1 - \frac{1}{2} \tau_{\sigma}^{\alpha} ( i \omega)^{\alpha}\right].
\label{Eq:lowFreq2}
\end{align}

Thus in the low frequency region it makes no difference if one deals with the simpler fractional Kelvin-Voigt wave equation. The resulting wave number is on form $\beta_k-i\alpha_k$ and under the condition of Eq. (\ref{eq:lowfreqCond}), the attenuation term, $\alpha_k = \alpha_0 \omega^{\alpha+1} = \alpha_0 \omega^y$,  is (\cite{Holm2010}):
\begin{equation}
\alpha_0 = \frac{\tau_\sigma^{\alpha}}{2c_0}\sin\frac{\pi \alpha}{2} = -\frac{\tau_\sigma^{y-1}}{2c_0}\cos\frac{\pi y}{2},
\label{Eq:low-alpha}
\end{equation}
where it also has been written in terms of the power law of the attenuation, i.e. $y=\alpha+1$. The low frequency condition of Eq. (\ref{eq:lowfreqCond}) combined with Eq. (\ref{Eq:low-alpha}) can then also be written as $2\alpha_0 c_0/\cos(\pi y/2) \ll 1$.


Recalling that the real part of the wave number $\beta_k-i\alpha_k$  has the subscript $k$ in order to distinguish it from the fractional order, the dispersion term is: 
\begin{align}
\beta_k = \frac{\omega}{c_0} \left(1-\frac{(\omega\taus)^{\alpha}}{2}\cos\frac{\pi \alpha}{2}\right) 
= \frac{\omega}{c_0} - \frac{\taus^{y-1}}{2c_0}\sin{(\frac{\pi y}{2})} \omega^y
\notag\\
= \frac{\omega}{c_0} + \alpha_0 \tan{(\frac{\pi y}{2})} \omega^y.
\label{Eq:low-beta}
\end{align}

In a similar manner the wave number for low frequencies may also be expressed with $y$ and $\alpha_0$, from Eq.\ (\ref{Eq:lowFreq2}):
\begin{align}
k \approx  \beta_k-i\alpha_k  
= \frac{\omega}{c_0} - \frac{\alpha_0 i^{y+1} \omega^y}{\cos{(\pi y/2)}}.
\label{Eq:lowFreq3}
\end{align}


A somewhat different approach to the low frequency case is to take the exact dispersion relation, Eq. (\ref{Eq:dispersionZener}), with $\tau_{\epsilon} = 0$, i.e.:
\begin{align}
k^2 - \frac{\omega^2}{c^2_0}  + (\tau_{\sigma}i \omega)^{\alpha} k^2  = 0
\label{Eq:dispersionKV}
\end{align}
It can be rewritten with a real and imaginary part for the loss term using the relation $i^\alpha = \cos(\pi\alpha/2) + i \sin(\pi\alpha/2)$:
\begin{align}
k^2 - \frac{\omega^2}{c^2_0} +   i \sin(\pi\alpha/2) (\tau_{\sigma} \omega)^{\alpha} k^2 + \cos(\pi\alpha/2) (\tau_{\sigma} \omega)^{\alpha}  k^2 = 0.
\label{Eq:dispersionLow}
\end{align}
The loss term has now been expanded into an imaginary part which mainly describes attenuation and a real part which relates to dispersion.

Because of the low frequency assumption,  Eq.\ (\ref{eq:lowfreqCond}), the loss term (the second term in Eq.\ (\ref{Eq:lowFreq2}) or the two last terms in Eq.\ (\ref{Eq:dispersionLow})), is much smaller than the first terms. This leads to:
\begin{align}
k \approx  \frac{\omega}{c_0}.
\label{Eq:lowlowFreq}
\end{align}

Therefore the wave number may further be approximated by substituting temporal differentation for spatial differentiation in that small loss term.


\pagebreak

%
%
\newcommand{\dtwoudttwo}{\ensuremath{\dfrac{\partial^2 u}{\partial t^2}}}
\newcommand{\dtwoudxtwo}{\ensuremath{\dfrac{\partial^2 u}{\partial x^2}}}
\newcommand{\dtwopdttwo}{\ensuremath{\dfrac{\partial^2 p}{\partial t^2}}}
\newcommand{\dtwopdxtwo}{\ensuremath{\dfrac{\partial^2 p}{\partial x^2}}}
\newcommand{\widehatbox}{\ensuremath{{\widehat\Box}}}

\end{document}